\documentclass[9pt,sigconf,nonacm]{acmart}
\usepackage{amsmath,amsfonts}
\usepackage{multirow}
\usepackage{multicol}
\usepackage{amsthm}
\usepackage{listings}
\usepackage{algorithm}
\usepackage{algpseudocode}
\usepackage{graphicx}
\usepackage{textcomp}
\usepackage{xcolor}
\usepackage{cleveref}
\usepackage{booktabs}
\usepackage{comment}
\usepackage{tabularx}
\usepackage{multirow}
\usepackage{bm}
\usepackage{url}
\usepackage{xspace}
\usepackage{pifont}
\usepackage{arydshln}
\usepackage{verbatim}
\usepackage[referable]{threeparttablex}
\usepackage{calc} 
\usepackage{color}
\usepackage{float}
\usepackage{subcaption}
\usepackage[utf8]{inputenc}
\usepackage{booktabs} % For prettier tables
\usepackage{tabularx} % For tables with adjustable-width columns
\usepackage{makecell}
\usepackage{tikz}
\usetikzlibrary{shapes.geometric, arrows}
\tikzstyle{startstop} = [rectangle, rounded corners, minimum width=3cm, minimum height=1cm,text centered, draw=black]
\tikzstyle{io} = [trapezium, trapezium left angle=70, trapezium right angle=110, minimum width=3cm, minimum height=1cm, text centered, draw=black]
\tikzstyle{process} = [rectangle, minimum width=3cm, minimum height=1cm, text centered, draw=black]
\tikzstyle{arrow} = [thick,->,>=stealth]

\usepackage{filecontents}                                  % support to pgfplots
\usepackage{pgfplots}
\usepackage{pgfplotstable}
\usepackage{scalefnt}
\pgfplotsset{compat=newest}

\definecolor{USTgold}{RGB}{153,102,0}
\definecolor{USTyellow}{RGB}{204,153,0}
\definecolor{USTyellowlight}{RGB}{255,212,0}
\definecolor{USTorange}{RGB}{255,166,26}
\definecolor{USTpink}{RGB}{255,157,157}
\definecolor{USTblue}{RGB}{0,51,102}
\definecolor{USTmiddle}{RGB}{0,116,188}
\definecolor{USTlight}{RGB}{99,202,225}
\definecolor{USTgray}{RGB}{204,204,204}
\definecolor{USTred}{RGB}{237,27,47}
\definecolor{USTdarkred}{RGB}{124,35,72}

\definecolor{CUHKorange}{RGB}{244,106,18} %F47012
\definecolor{CUHKblue}{RGB}{0,111,190}    %006FBE
\definecolor{CUHKgreen}{RGB}{0,127,128}   %007F80
\definecolor{CUHKred}{RGB}{228,46,36}     %E42E24
\definecolor{CUHKyellow}{RGB}{198,148,34} %C69422
\definecolor{CUHKdark}{RGB}{114,44,114}   %722C72
\definecolor{CUHKmiddle}{RGB}{144,44,144} %902C90
\definecolor{CUHKlight}{RGB}{167,44,167}

\iftrue
\def\BibTeX{{\rm B\kern-.05em{\sc i\kern-.025em b}\kern-.08em
    T\kern-.1667em\lower.7ex\hbox{E}\kern-.125emX}}

\fi

\iftrue
\fi

\iftrue
\usepackage{titlesec}
\titlespacing\section{2pt}{5pt plus 1pt minus 1pt}{0pt plus 1pt minus 1pt}
\titlespacing\subsection{2pt}{5pt plus 1pt minus 1pt}{0pt plus 1pt minus 1pt}
\titlespacing\subsubsection{2pt}{5pt plus 1pt minus 1pt}{2pt plus 1pt minus 1pt}
\usepackage[inline]{enumitem}
\setlist{leftmargin=5.08mm}
\fi

\iftrue
\fi

\newcommand{\cmark}{\raisebox{0.12ex}{\scalebox{1.35}{\ding{51}}}}

\def\pmark{\tikz\draw[scale=0.4,fill=black](0,.35) -- (.25,0) -- (1,.7) -- (.25,.15) -- cycle (0.75,0.2) -- (0.77,0.2)  -- (0.6,0.7) -- cycle;}
\usepackage[table]{xcolor}
\definecolor{lightred}{RGB}{255,220,220}

\begin{document}
\title{
AssertLLM2: A Comprehensive LLM Benchmark for Assertion Generation from Design Specifications
}

% \iffalse
% \author{
% \IEEEauthorblockN{Yuchao Wu}
% \IEEEauthorblockA{HKUST}
% \and
% \IEEEauthorblockN{Zhiyao Xie}
% \IEEEauthorblockA{HKUST}
% }
% \fi
\author{Yuchao Wu$^{\dagger}$, Wenji Fang$^{\dagger}$, Jing Wang, Wenkai Li, Ziyan Guo, Zhiyao Xie$^{\ast}$}
\affiliation{Hong Kong University of Science and Technology (HKUST)}

\thanks{$^{\dagger}$ Equal Contribution. $^{\ast}$ Corresponding Author (eezhiyao@ust.hk)}

\begin{abstract}
Assertion-based verification (ABV) is a cornerstone of modern hardware design, yet manually translating design intent into formal SystemVerilog Assertions (SVAs) remains labor-intensive and error-prone. 
While Large Language Models (LLMs) show promise for automating this process, existing benchmarks remain limited by unrealistic task formulations, weak specification inputs, and oversimplified evaluation. 
To address these limitations, we introduce \textbf{AssertLLM2}, an open-source benchmark for realistic assertion generation in hardware verification. 
AssertLLM2 contains 83 real-world designs across 13 functional categories. 
For each design, the benchmark provides a structured design specification, a verified dependency-complete golden RTL, and systematically mutated buggy RTL variants. 
These support two practical settings: \emph{bug-prevention}, where assertions are generated from specifications to guard against design errors, and \emph{bug-hunting}, where assertions are generated to expose discrepancies between intended behavior and faulty implementations. 
To the best of our knowledge, AssertLLM2 is the first benchmark to explicitly use buggy RTL as input to evaluate bug-detection capability. 
AssertLLM2 further adopts a more rigorous evaluation framework spanning syntactic validity, formal provability, coverage, and mutation-based bug detection. 
Our benchmark enables a more realistic and extensive assessment of assertion generation and establishes rigorous baselines for state-of-the-art LLMs in practical hardware verification.
\end{abstract}
\maketitle
\pagestyle{empty}

 \section{Introduction}
\label{sec:intro}

Verification plays a critical role in ensuring the reliability, controlling the development costs, and reducing the time-to-market of modern chip designs \cite{liu2026llm}. 
Unlike software defects, which can often be corrected after deployment, post-silicon hardware bugs are extraordinarily expensive to diagnose and fix, and may even result in product recalls. 
During verification, assertion-based verification (ABV) is widely adopted as it represents expected design behavior using SystemVerilog Assertions (SVAs), thereby enabling precise, machine-checkable analysis and more effective verification \cite{witharana2022survey}.

Despite its practical effectiveness, the generation of high-quality assertions remains a major challenge. 
In industrial verification flows, engineers have to manually translate high-level specifications into formal properties that are both semantically sound and consistent with low-level RTL behavior \cite{foster2008assertion}. 
This process demands substantial expertise in specification interpretation, signal semantics, timing relationships, and design structure. As hardware systems continue to increase in complexity, assertion generation has become increasingly labor-intensive, error-prone, and reliant on expert knowledge, making it a major bottleneck in contemporary hardware verification.

Recent advances in Large Language Models (LLMs) have shown strong potential for generating SVAs. 
To bridge the semantic gap between high-level specifications and verification properties, recent works have explored LLM-assisted assertion generation by enhancing specification grounding \cite{ wu2025spec2assertion, fu2026chatsva, tian2025assertcoder}, incorporating richer RTL context \cite{bai2025assertionforge, lyu2025assertgen, lyu2026assertminer}, introducing structured refinement procedures \cite{lyu2025assertfix, mali2024chiraag}, and task-specific fine-tuning \cite{shahidzadeh2025automated, thangellamudi2026bridging}.

\begin{figure}[!t]
  \centering
  \includegraphics[width=0.9\linewidth]{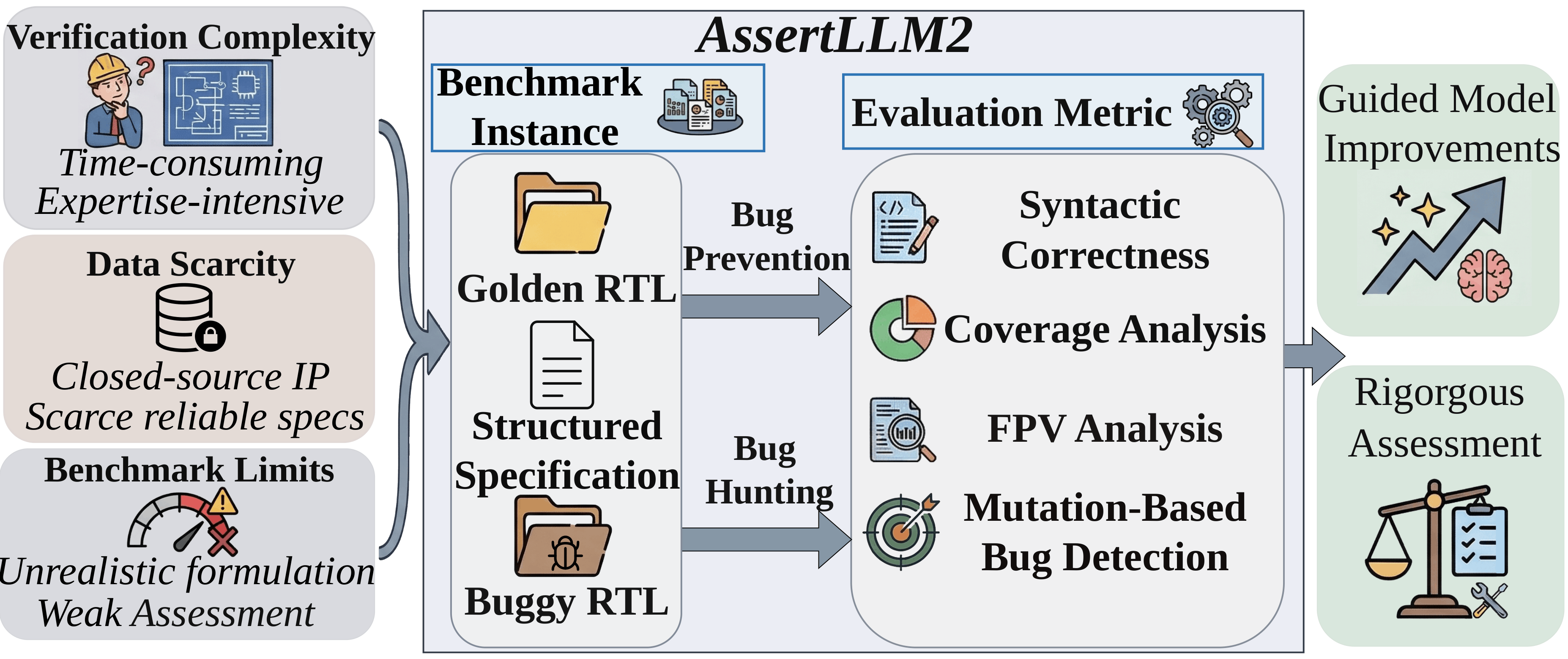}
  \vspace{-0.3cm}
  \caption{Overview of the AssertLLM2 framework.
  Unlike prior benchmarks, AssertLLM2 adopts a more practical problem formulation, supports both bug-prevention and bug-hunting settings, provides richer benchmark instances, and enables more rigorous and faithful evaluation to better assess and guide LLM improvement. } 
  \label{fig:framework}
\end{figure}

\begin{table*}[t]
\centering
\resizebox{0.98\textwidth}{!}{
\begin{tabular}{l c c c c c c c c c c c c} 
\toprule
\multirow{3}{*}{\textbf{Works}} 
& \multicolumn{2}{c}{\textbf{Task}} 
& \multirow{3}{*}{\textbf{\makecell[c]{SPEC\\Format}}} 
& \multicolumn{3}{c}{\textbf{Dataset}} 
& \multicolumn{6}{c}{\textbf{Evaluation Metric}} \\
\cmidrule(lr){2-3} \cmidrule(lr){5-7} \cmidrule(lr){8-13}

& \multirow{2}{*}{\textbf{\makecell[c]{Bug\\Preventing}}}
& \multirow{2}{*}{\textbf{\makecell[c]{Bug\\Hunting}}}
& 
& \multirow{2}{*}{\textbf{\# Designs}} 
& \multirow{2}{*}{\textbf{\shortstack[c]{Sample\\Scale}}} 
& \multirow{2}{*}{\textbf{\makecell[c]{Avg.\\LOC.}}} 
& \multirow{2}{*}{\textbf{Syntax}} 
& \multicolumn{3}{c}{\textbf{Coverage}} 
& \multirow{2}{*}{\textbf{\makecell[c]{FPV\\Result}}} 
& \multirow{2}{*}{\textbf{\makecell[c]{Mutation-Based \\ Bug Detection}}} \\
\cmidrule(lr){9-11}

& & & 
& & & 
& 
& \textbf{COI} 
& \textbf{Proof} 
& \textbf{Formal} 
& & \\
\midrule

VERT\cite{menon2025vert} 
& --
& $\pmark ^ \ast $
& --
& 20000 
& Snippet 
& 18
& \cmark 
& \cmark
& --
& --
& \cmark 
& \cmark \\ \addlinespace

AssertionBench\cite{pulavarthi2025assertionbench}
& --
& $\pmark ^ \ast $
& --
& 101 
& Module 
& 123
& \cmark 
& --
& --
& -- 
& \cmark 
& -- \\ \addlinespace

FVEval\cite{kang2025fveval}
& \cmark
& $\pmark ^ \ast $  
& {\makecell[c]{Partial \\ Sentences}}
& 379 
& Snippet 
& 73
& \cmark
& --
& --
& -- 
& \cmark 
& -- \\ \addlinespace 

CVDP$^\star$ \cite{pinckney2025cvdp} 
& --
& $\pmark ^ \ast $ 
& \makecell[c]{Pre-extracted \\ Checklists}
& \makecell[c]{98} 
& Module/IP 
& 167
& \cmark 
& --
& --
& --
& -- 
& -- \\ \addlinespace

AssertLLM\cite{fang2024assertllm}
& \cmark
& -- 
& \makecell[c]{Unstructured \\ Raw PDFs}
& 20 
& IP/System
& 5146 
& \cmark 
& \cmark
& --
& --
& \cmark 
& -- \\ \midrule 

\textbf{Ours} 
& \cmark
& \cmark 
& \textbf{\makecell[c]{Structured \& \\ Unified SPEC}} 
& \textbf{83}
& \textbf{IP/System} 
& \textbf{2888}
& \cmark 
& \cmark 
& \cmark
& \cmark
& \cmark 
& \cmark \\

\bottomrule
\end{tabular}
}
\begin{tablenotes}\footnotesize
\item $^\star$ CVDP~\cite{pinckney2025cvdp} is a comprehensive benchmark; the data reported herein solely reflects its assertion-specific subset (category \texttt{c014}).
\item $^\ast$ Marked works use the golden RTL during assertion generation, rather than the buggy RTL under verification.
\end{tablenotes} 
\caption{Comparison of assertion generation benchmarks. AssertLLM2 enables a more practical problem formulation by integrating \textit{bug-prevention} and \textit{bug-hunting}, providing a structured specification, and establishing a rigorous evaluation methodology based on FPV results, comprehensive coverage analysis, and mutation-based bug detection on complex designs.}
\label{tab:assert_benchmark_compare}
\end{table*}

Despite these advances in methodologies, rigorous evaluation remains a major challenge. 
Benchmarks are vitally important since they directly determine how models are evaluated. 
However, public hardware data and hardware verification benchmarks remain scarce. 
Furthermore, existing benchmarks about LLM-assisted assertion generation \cite{menon2025vert, pulavarthi2025assertionbench, kang2025fveval,pinckney2025cvdp,fang2024assertllm} still suffer from several fundamental limitations. 
As summarized in \Cref{tab:assert_benchmark_compare}, these limitations fall into three main aspects.

\textbf{Limitation 1: Unrealistic formulation.} 
Some existing benchmarks \cite{menon2025vert, pulavarthi2025assertionbench, kang2025fveval, pinckney2025cvdp} rely on golden RTL to construct and evaluate assertion generation tasks. 
This formulation is fundamentally unrealistic, since golden RTL is not available in real verification scenarios. 
Assertions are used to help verify a design before correctness is established, not to be derived from an already verified implementation. 
As a result, these benchmarks reduce assertion generation to a translation of golden RTL, rather than evaluating whether a model can generate useful assertions for realistic verification tasks.
    
\textbf{Limitation 2: Oversimplified evaluation.}
Prior benchmarks \cite{menon2025vert, pulavarthi2025assertionbench, kang2025fveval,pinckney2025cvdp,fang2024assertllm} use severely insufficient evaluation metrics, which leads to an incomplete evaluation of whether generated assertions are truly useful in practical verification.
As shown in Table~\ref{tab:assert_benchmark_compare}, existing benchmarks largely focus on syntax and FPV results when evaluating generated assertions.
This means that existing benchmarks are effective for checking whether assertions can be parsed and proven, but they can not determine whether those assertions constrain relevant logic, improve meaningful formal coverage, or detect design errors.
In other words, assertions that are syntactically correct and formally provable may still be trivial, vacuous, or weak in practical verification.
Consequently, relying on existing benchmarks can overestimate both assertion quality and the practical utility of LLM-generated assertions.

\textbf{Limitation 3: Limited design scale and specification quality.}
Existing specification-driven benchmarks \cite{kang2025fveval,pinckney2025cvdp,fang2024assertllm} remain limited in two key aspects: design coverage and specification input. 
For design coverage, Most prior benchmarks \cite{kang2025fveval,pinckney2025cvdp} focus on small module-level designs, rather than the larger and more complex IP-level designs encountered in practice. 
The only benchmark targeting a larger-scale setting \cite{fang2024assertllm} still includes fewer than 20 designs and offers limited diversity.
This makes existing benchmarks insufficient for evaluating whether the models can generalize across realistic hardware designs.
For specification quality, prior work often relies on either simplified requirement inputs \cite{pinckney2025cvdp,kang2025fveval} or raw documents \cite{fang2024assertllm} with inconsistent quality and structure. 
Simplified checklists reduce the difficulty of generating assertions from realistic design intent, while raw documents make evaluation depend not only on assertion generation, but also on document quality and multimodal information spread across text, tables, and figures \cite{alshazly2014detecting, chang2024natural}.
Consequently, prior benchmarks fall short of providing a realistic and comprehensive evaluation of practical assertion generation.

% FVP result or something..
% Use AsserLLMv2
% Coverage expand, emphasize

% Too short, align with limitation (1, 2, 3)
To address these limitations, we introduce \textbf{AssertLLM2\footnote{Our benchmark is open-sourced in \url{https://github.com/hkust-zhiyao/AssertLLM2}, including the dataset, evaluation scripts, LLM prompts, and buggy-RTL generation scripts.}}, an open-source benchmark comprising 83 real-world system-level designs for realistic SVA generation. 
Specifically, AssertLLM2 makes the following three contributions:

\ding{182} \textbf{More complete and realistic task formulation.} 
Compared with prior benchmarks that are either RTL-centric or confined to the $SPEC \rightarrow SVA$ setting, AssertLLM2 supports two complementary assertion-generation scenarios corresponding to two practical stages of hardware development.
The first is \emph{bug-prevention}. 
This scenario reflects the stage where RTL is still under development. Assertions are generated from the specification alone to capture intended behavior and help prevent design errors early.
The second is \emph{bug-hunting}. 
This scenario reflects the stage where RTL has been implemented but is not yet fully verified, and may therefore still contain functional bugs. 
Assertions are generated from the specification together with the RTL to test whether they can expose mismatches between intended behavior and implementation behavior. To the best of our knowledge, 
AssertLLM2 is the first benchmark to explicitly use buggy RTL as input to evaluate bug-detection capability.
By supporting both settings within a unified benchmark, AssertLLM2 captures two principal practical roles of assertions: preserving design intent during development, and exposing functional errors during verification. 
In both settings, the golden RTL is used only as an evaluation oracle, rather than as generation input. 

\ding{183} \textbf{More rigorous evaluation.}
AssertLLM2 adopts a unified evaluation framework for assessing generated assertions from multiple complementary perspectives. 
Beyond syntax and FPV, it evaluates COI coverage, proof coverage, formal coverage, and mutation-based bug detection. 
Together, these metrics capture whether assertions constrain relevant logic, improve meaningful formal coverage, and expose design errors, rather than merely passing surficial correctness checks.
This allows AssertLLM2 to distinguish assertions that are only superficially valid from those that provide real verification value. 
As a result, the benchmark supports a more rigorous and practically informative assessment of assertion quality.

% reduce spec and add data
\ding{184} \textbf{Richer and more realistic designs and specifications.}
AssertLLM2 addresses the limitations of prior benchmarks in design scale and diversity by including 83 real-world system-level designs spanning diverse functionalities and architectures.
For each design, the benchmark includes a structured textual specification, the raw PDF, a verified golden RTL reference, and systematically mutated buggy RTL variants. 
The structured specification removes unrelated content, converts figure information into text, and improves lower-quality specifications into a clearer representation of design intent, which helps mitigate the limitations of specification quality.
Together, these components establish a unified and realistic benchmark for assertion generation and evaluation.

Using AssertLLM2, we further evaluate state-of-the-art LLMs to establish strong baselines for specification-driven assertion generation and to reveal their capabilities and limitations in practical verification settings.

% Our key contributions are as follows:
% \begin{itemize}
%     \item We introduce \textbf{AssertLLM2}, an open-source benchmark comprising 83 real-world system-level designs. 
%     Each benchmark instance provides three aligned, high-quality artifacts: a uniformly formatted design specification, a verified golden RTL oracle, and comprehensively mutated buggy RTL variants.

%     \item We construct systematically mutated buggy RTL variants for each design, instantiating the bug-hunting setting and enabling downstream evaluation of assertion-based fault detection.

%     \item We develop a comprehensive open-source evaluation framework that jointly assesses syntactic validity, formal provability, structural coverage, and mutation-based fault detection, going beyond the insufficient metric settings adopted in prior work.

%     \item We evaluate state-of-the-art LLMs on AssertLLM2, establishing rigorous baselines for specification-driven assertion generation and revealing their capabilities and limitations in practical verification settings.
% \end{itemize}

\section{Problem Formulation}

% % Emphasize the previous work's problem
% In this section, we provide a general formulation of the assertion generation task. Let $S$ denote the natural language design specification, $R_{\text{bug}}$ denote a buggy RTL implementation, and $\mathcal{A}=\{a_1,\dots,a_N\}$ denote the generated set of SVAs. 
% We use $\mathcal{f}(\cdot)$ to denote the abstract generation mapping from the available task inputs to the assertion set produced by a model. 
% To reflect practical verification usage, we consider two complementary SVA-generation paradigms, distinguished by the artifacts exposed to the generator.

% % Task 1
% \paragraph{\textbf{Problem 1: bug-prevention.}}
% The generator is given only the $S$ and must produce $\mathcal{A}$ without access to implementation code:
% \begin{equation}
% \mathcal{A} = f_{prevent}(S).
% \end{equation}
% The generated $\mathcal{A}$ is then evaluated downstream to assess syntactic validity, formal soundness, and verification effectiveness.

% % Task 2
% \paragraph{\textbf{Problem 2:  bug-hunting.}}
% The generator is given both the $S$ and a buggy RTL implementation $R_{\text{bug}}$, and produce $\mathcal{A}$ conditioned on both:
% \begin{equation}
% \mathcal{A} = f_{hunt}(S, R_{\text{bug}}).
% \end{equation}
% The generated $\mathcal{A}$ is then evaluated downstream for its ability to expose discrepancies in faulty implementations.

Let $S$ denote the natural-language design specification, \(R_{\text{bug}}\) denote the buggy RTL implementation, and \(R_{\text{gold}}\) denote the verified golden RTL.
We use $\mathcal{F}(\cdot)$ to represent the LLM generation process, which maps the provided input context to a set of generated SVAs $\mathcal{A}$.

\paragraph{\textbf{Problem 1: bug-prevention.}}
In the \textit{bug-prevention} setting, assertions are generated from specifications to guard against design errors.
The generator is given only the specification $S$ and must produce $\mathcal{A}$ without access to implementation code:
\begin{equation}
\mathcal{A} = \mathcal{F}_{\text{prevent}}(S).
\end{equation}
The generated $\mathcal{A}$ is then evaluated in our evaluation framework.
We denote the evaluation mapping for this setting as $\mathcal{E}_{\text{prevent}}$, which computes an evaluation vector $\mathcal{V}$:
\begin{equation}
\mathcal{V} = \mathcal{E}_{\text{prevent}}(\mathcal{A}, R_{\text{gold}}, R_{\text{bug}}),
\end{equation}
where $\mathcal{V}$ consists of syntax correctness, FPV results, mutation-based bug detection, and coverage metrics including COI coverage, proof coverage, and formal coverage.

\paragraph{\textbf{Problem 2: bug-hunting.}} 
In the \textit{bug-hunting} setting, assertions are generated to expose discrepancies between intended behavior and faulty implementations.
The generator is provided with both the specification $S$ and a buggy RTL implementation $R_{\text{bug}}$:
\begin{equation}
\mathcal{A} = \mathcal{F}_{\text{hunt}}(S, R_{\text{bug}}).
\end{equation}
The generated assertions are evaluated under the corresponding benchmark framework:
\begin{equation}
\mathcal{V} = \mathcal{E}_{\text{hunt}}(\mathcal{A}, R_{\text{gold}}, R_{\text{bug}}),
\end{equation}
where $\mathcal{V}$ mainly focuses on bug kill ratio.

\section{Benchmark Data Overview}

AssertLLM2 comprises 83 real-world designs spanning 13 functional categories. 
As summarized in Table~\ref{tab:dataset_design_summary}, the designs are organized into representative groups within each category, based on shared functionality and architectural characteristics. 
These 13 categories and altogether 37 groups encompass a wide range of design types and architectural styles, with representative examples such as CPU cores, signal-processing pipelines, cryptographic cores, and memory controllers, reflecting the substantial diversity of AssertLLM2.
For each group, the table reports: a brief functional description, the average RTL size in lines of code, the raw and structured specification token counts, and the number of included designs.

\begin{figure}[!t]
\centering
\includegraphics[width=\linewidth]{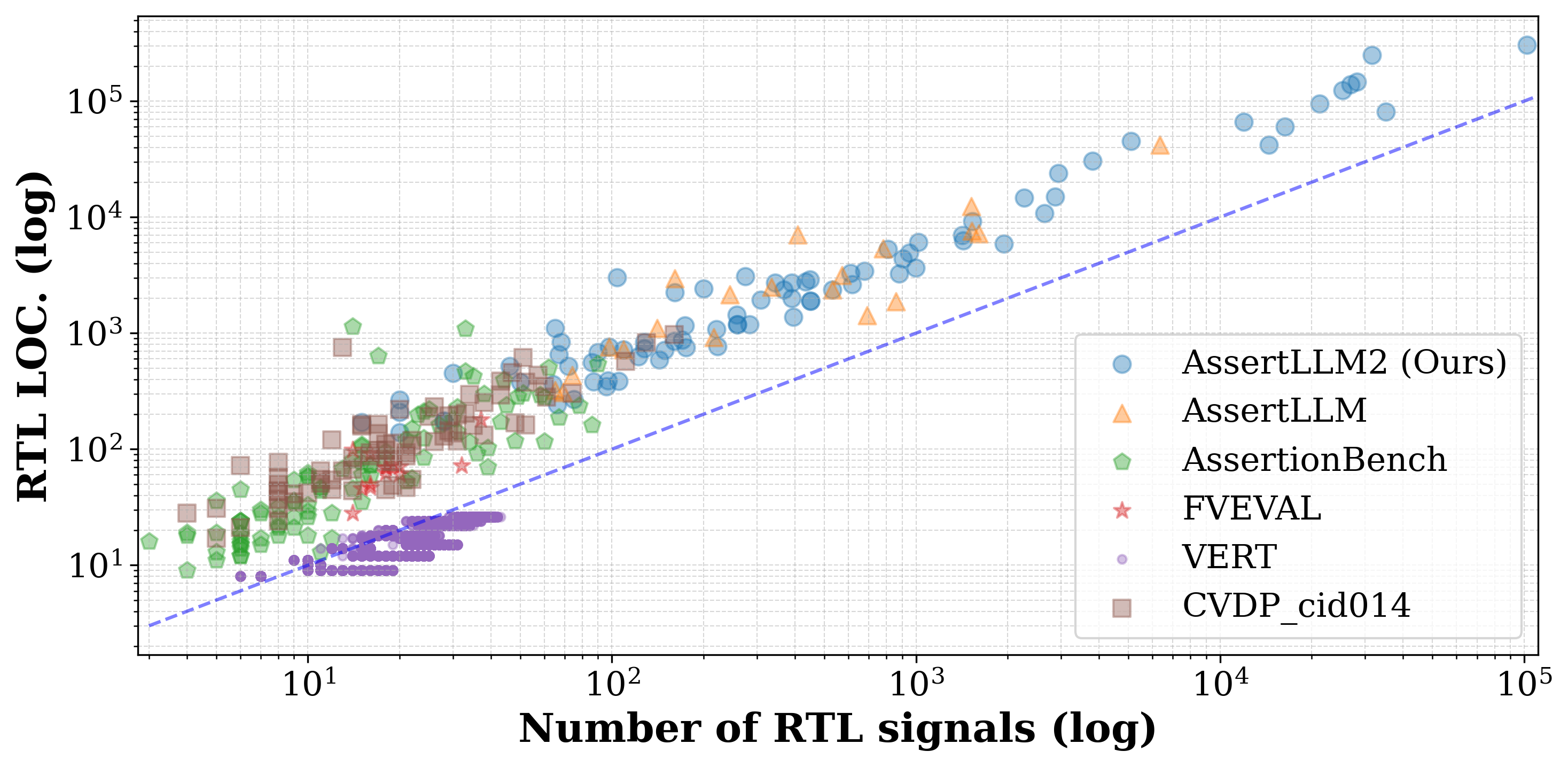}
\vspace{-0.8cm}
\caption{Comparison of design-level scale (in both RTL lines and number of signals) across public assertion-generation benchmarks, where each point denotes a benchmark instance.}
\label{fig:design_scale_compare}
\end{figure}

Beyond this broad diversity, AssertLLM2 also stands out from prior benchmarks in the scale of its designs.
Prior assertion-generation benchmarks are often built around small modules, simplified instances, or snippet-level samples \cite{menon2025vert, pulavarthi2025assertionbench, kang2025fveval, pinckney2025cvdp, fang2024assertllm}, whereas AssertLLM2 is constructed from complete system-level designs. 
As shown in \Cref{fig:design_scale_compare}, AssertLLM2 is compared against prior benchmarks using RTL lines of code and RTL signal count, which capture implementation size and design complexity, respectively. 
Under both measures, AssertLLM2 consists of substantially larger designs, making it more representative of realistic assertion generation scenarios.

For each design included in AssertLLM2, we provide three aligned components: (1) a structured specification\footnote{We also retain the raw specification PDF as the original reference document, and users can choose either our structured specification or the raw PDF as input.}, (2) a golden reference RTL implementation, and (3) buggy RTL variants derived from the golden RTL with mutation-based techniques.
The buggy RTL variants include 20 single-bug variants and one five-bug variant obtained by combining five single-bug variants. 
The structured specification serves as the primary input to assertion generation, while the golden RTL implementation is used only for evaluation.
In the \textit{bug-prevention} setting, the buggy RTL variants are used only for evaluation. 
In the \textit{bug-hunting} setting, the five-bug variant is provided as input to better approximate a realistic faulty design, whereas the corresponding single-bug variants are reserved for evaluation so that each injected fault can be assessed independently.

\begin{table*}[t]
\centering
\resizebox{0.95\textwidth}{!}{
\begin{tabular}{l l p{6.5cm} c c c c}
\toprule
\multirow{2}{*}{\textbf{Category}} & \multirow{2}{*}{\textbf{Group}} & \multirow{2}{*}{\textbf{Brief Description}} & \multirow{2}{*}{\textbf{Avg. LOC}} & \multicolumn{2}{c}{\textbf{SPEC Tokens}} & \multirow{2}{*}{\textbf{\# Designs}} \\
\cmidrule(lr){5-6}
 & & & & \textbf{raw} & \textbf{ours} & \\
\midrule
\multirow{4}{*}{Arithmetic Core}
& arithmetic datapath & Datapath designs for arithmetic computation. & 619 & 3798 & 2464 & 2 \\
& trigonometric FPU & Hardware trigonometric computation unit. & 838 & 5319 & 2068 & 1 \\
& graphics accelerator & Arithmetic engine for graphics workloads. & 3267 & 30059 & 4455 & 1 \\
& arithmetic utilities & Counter, noise, and decoding arithmetic logic. & 571 & 3136 & 1861 & 3 \\
\midrule

\multirow{8}{*}{\makecell{Communication \\ Controller}}
& Ethernet MAC & Ethernet frame transmit/receive MAC. & 3716 & 9048 & 4710 & 4 \\
& i\textsuperscript{2}c & I\textsuperscript{2}C serial bus controllers and host bridges. & 1250 & 12325 & 3599 & 4 \\
& IEEE 1588 & Precision timestamping and clock-sync logic. & 1386 & 618 & 3139 & 1 \\
& SPI & SPI flash, master, and protocol-bridge controllers. & 1282 & 11832 & 3067 & 5 \\
& UART & UART peripherals and UART-centric bus bridges. & 605 & 4017 & 2654 & 4 \\
& SATA phy & SATA physical-layer interface logic. & 2727 & 1362 & 4065 & 1 \\
& SDMMC host controller & SD/MMC command and data host controllers. & 2788 & 73377 & 5530 & 3 \\
& sport interface & Synchronous serial port controller. & 588 & 1079 & 3180 & 1 \\
\midrule

Coprocessor
& xgate & Auxiliary task-offload coprocessor core. & 2886 & 40164 & 4799 & 1 \\
\midrule

\multirow{3}{*}{Crypto Core}
& AES & AES-family symmetric encryption engine. & 796 & 2777 & 2060 & 4 \\
& pairing-based crypto core & Pairing-oriented cryptographic computation core. & 1486 & 2952 & 2165 & 2 \\
& lightweight security core & Lightweight ciphers and hash accelerators. & 421 & 2015 & 1714 & 3 \\
\midrule

\multirow{4}{*}{DSP Core}
& IMA adpcm codec & IMA ADPCM audio codec engine. & 237 & 3582 & 1418 & 2 \\
& PID controller & Digital proportional-integral-derivative controller. & 1935 & 3041 & 2547 & 1 \\
& FFT processor & Fast Fourier transform processor. & 753 & 2900 & 2501 & 1 \\
& ECG signal chain & ECG acquisition, filtering, and extraction datapath. & 1422 & 2387 & 2483 & 1 \\
\midrule

\multirow{2}{*}{ECC Core}
& reed-solomon decoder & Reed--Solomon error-correction decoder. & 3305 & 1990 & 3154 & 1 \\
& gf256 rs core & GF(256)-based Reed--Solomon codec core. & 453 & 1365 & 2018 & 1 \\
\midrule

Library
& utility library & Reusable RTL utility components. & 381 & 4712 & 2254 & 2 \\
\midrule

\multirow{2}{*}{Memory Core}
& SDram controller & External SDRAM access controller. & 2280 & 7162 & 3797 & 3 \\
& memory buffer & On-chip RAM and FIFO buffering logic. & 1189 & 3439 & 2110 & 2 \\
\midrule

\multirow{5}{*}{Processor}
& embedded CPU & Embedded general-purpose processor cores. & 3415 & 5086 & 2423 & 4 \\
& RISC CPU & Modern embedded RISC processor core. & 10312 & 50676 & 10552 & 2 \\
& legacy-compatible CPU & Legacy ISA-compatible processor cores. & 19610 & 11736 & 2850 & 4 \\
& educational CPU & Teaching-oriented simplified processor core. & 627 & 3658 & 2742 & 1 \\
& graphics processing unit & Parallel graphics compute engine. & 6068 & 13869 & 2975 & 1 \\
\midrule

System Controller
& system controller family & Memory controller and PCI bridge logic. & 9225 & 30876 & 5152 & 2 \\
\midrule

System-on-Chip
& Wishbone SOC IP family & Wishbone SoC interconnect and DMA backbone. & 6438 & 8008 & 3468 & 2 \\
\midrule

\multirow{2}{*}{Video Controller}
& display compositor & Text and sprite display compositor. & 962 & 1914 & 3580 & 2 \\
& video display pipeline & Video timing and pixel output pipeline. & 1602 & 12995 & 3236 & 2 \\
\midrule

\multirow{3}{*}{Other}
& AC97 controller & AC'97 audio controller. & 1897 & 6949 & 3584 & 2 \\
& utility modules & Miscellaneous low-complexity support RTL. & 372 & 6359 & 2122 & 4 \\
& timebase controller & Timer and PWM generation peripherals. & 380 & 3514 & 2743 & 3 \\

\bottomrule
\end{tabular}
}
\caption{AssertLLM2 benchmark summary. Designs are organized by category and grouped into shared design families for concise presentation. For each group, we report a brief functional description, average RTL size (LOC), raw-spec token count, structured-spec token count, and the number of designs.}
\label{tab:dataset_design_summary}
\end{table*}

\section{Benchmark Data Construction}
\subsection{Data Collection}

% list the source/ summarize the range
We construct the dataset by automatically collecting open-source repositories \cite{freecores,e203_hbirdv2,opencores,vexriscv} that provide Verilog/SystemVerilog designs along with comprehensive specifications. 
As these specifications usually describe the behavior of the overall project or system, rather than that of a single module, we identify a top-level module for each repository so that the RTL can be interpreted as a complete design instance. 
We then apply a JasperGold \cite{jaspergold2015} elaboration check to verify that the selected top module can be elaborated as a complete and valid RTL design. 
Finally, we require at least 200 lines of RTL code to filter out trivial designs.

\begin{table*}[t]
\centering
\small
\renewcommand{\arraystretch}{1.12}
\caption{Mutation Operators used in buggy RTL generation.}
\vspace{-0.2 cm}
\begin{tabularx}{0.9\textwidth}{
    >{\raggedright\arraybackslash}p{0.10\textwidth}
    >{\raggedright\arraybackslash}p{0.26\textwidth}
    >{\raggedright\arraybackslash}X}
\hline
\textbf{Group} & \textbf{Operators} & \textbf{Effect} \\
\hline

\multirow{4}{0.20\textwidth}{Indexing and \\ bit selection}
& Index offset perturbation
& Mutates pointer or array index constants. \\
& Shift amount perturbation
& Mutates shift distances in shift expressions. \\
& Part select boundary perturbation
& Changes part select endpoints under guarded conditions. \\
& Slice endianness inversion
& Mirrors a part select slice within its declared range while preserving width. \\
\hline

\multirow{6}{0.20\textwidth}{Expression and \\ control semantics}
& Ternary branch mutation
& Forces the true/false branch, or swaps branches in a conditional expression. \\
& Logical relational operator replacement
& Replaces logical or relational operators according to predefined swap rules. \\
& Validation-guard forcing
& Forces a validity-related predicate in an \texttt{if} condition to constant true/false. \\
& Case semantics mutation
& Perturbs case selection behavior or case arm interpretation. \\
& If branch removal
& Deletes an if branch. \\
& Case branch removal
& Deletes a case branch. \\
\hline

\multirow{4}{0.20\textwidth}{Constants and \\ parameters}
& Statement constant perturbation
& Mutates integer constants inside assignment-rich procedural blocks. \\
& Assignment value constant perturbation
& Mutates constants on the right-hand side of assignments. \\
& Instance parameter perturbation
& Perturbs parameter arguments at module instantiation sites. \\
& Delay constant perturbation
& Mutates delay values in delay statements. \\
\hline

\multirow{3}{0.20\textwidth}{Connectivity and \\ dataflow}
& Named port connection swap
& Swaps two named port connections within an instance. \\
& Concatenation operand swap
& Swaps operands or components in concatenation expressions. \\
& Assignment duplication injection
& Inserts an additional mutated assignment into a procedural block. \\
\hline
\end{tabularx}
\label{tab:OP}
\end{table*}

\subsection{Structured Specification Construction}

In this subsection, we describe the construction of the structured specification.
Raw design specifications often vary substantially in quality and organization across projects. 
They often mix functional requirements with auxiliary content, such as implementation notes and revision history, and distribute important information across text, figures, and tables. 
This makes the documents difficult to use directly and creates a multimodal bottleneck for purely text-based open-source LLMs. 
Moreover, some raw documents are of relatively low quality and cannot be used directly as reliable inputs for benchmark construction.
To address these issues and establish a uniform evaluation baseline, we construct normalized, structured textual specifications from the raw documents.
We introduce the structured specification construction process below.

% Each structured specification follows a common schema that includes: module overview and design intent, top-level interface and signal semantics, architecturally visible state, functional behavior, verification-relevant disambiguating details, submodule and hierarchy context, and timing, reset, constraint, and corner-case notes. 
% Although different designs emphasize different aspects, the overall organization is kept consistent across the benchmark.

% The structured specification is produced through an LLM-assisted curation workflow. 
% Starting from the source documents, we first remove content that is not relevant to functional requirements, including revision histories, navigation text, repeated boilerplate, coding examples, and build environment configurations. 
% We then use an LLM to generate a normalized draft under a fixed section template. 
% This draft standardizes terminology, merges duplicated statements, converts figure semantics into text, and consolidates scattered requirement fragments into coherent descriptions. 
% For low-quality raw specifications, the same workflow is used to reorganize incomplete or unevenly presented information from the original documents into a clearer and more complete benchmark artifact. 
% The draft is subsequently reviewed against the source documentation at the clause level by human experts.
% The final specification is therefore a curated and human-validated benchmark artifact.

Each structured specification is produced through an LLM-assisted curation workflow with four main steps. 
\textbf{(1) Remove redundancy.} Starting from the source documents, we remove content that is not directly relevant to functional requirements, such as revision histories, navigation text, repeated boilerplate, coding examples, and build environment configurations. \textbf{(2) Generate a normalized draft.} We then use an LLM to organize the remaining content under a fixed section template. Each structured specification follows a common schema that includes module overview and design intent, top-level interface and signal semantics, architecturally visible state, functional behavior, verification-relevant disambiguating details, submodule and hierarchy context, and timing, reset, constraint, and corner-case notes. 
Although different designs emphasize different aspects, the overall organization is kept consistent across the benchmark.
\textbf{(3) Refine requirement descriptions.} 
The draft further standardizes terminology, merges duplicated statements, converts information conveyed in figures and tables into text, and consolidates requirement fragments scattered across the original documents into coherent descriptions. 
For lower-quality source documents, the same process is also used to reorganize incomplete or unevenly presented information into a clearer and more complete specification.
\textbf{(4) Validate against the source.} 
Finally, human experts review the draft against the original documentation at the clause level. 
The resulting structured specification is therefore a curated and human-validated benchmark artifact.

\subsection{Mutation-Based Buggy RTL Generation}
\label{subsec:mutate}
In addition to the structured specifications and golden RTL, we provide a curated buggy RTL corpus for each design. 
These variants are derived from the golden reference implementation through systematic mutation while preserving the original design context, top-level interface, and module hierarchy. 
Each buggy RTL is required to pass a JasperGold check and to differ functionally from the original RTL, which removes invalid or equivalent mutants (i.e., buggy RTL generated with mutation). 
We describe the buggy RTL generation process in detail below.

\begin{figure}[b]
\centering
\includegraphics[width=0.93\linewidth]{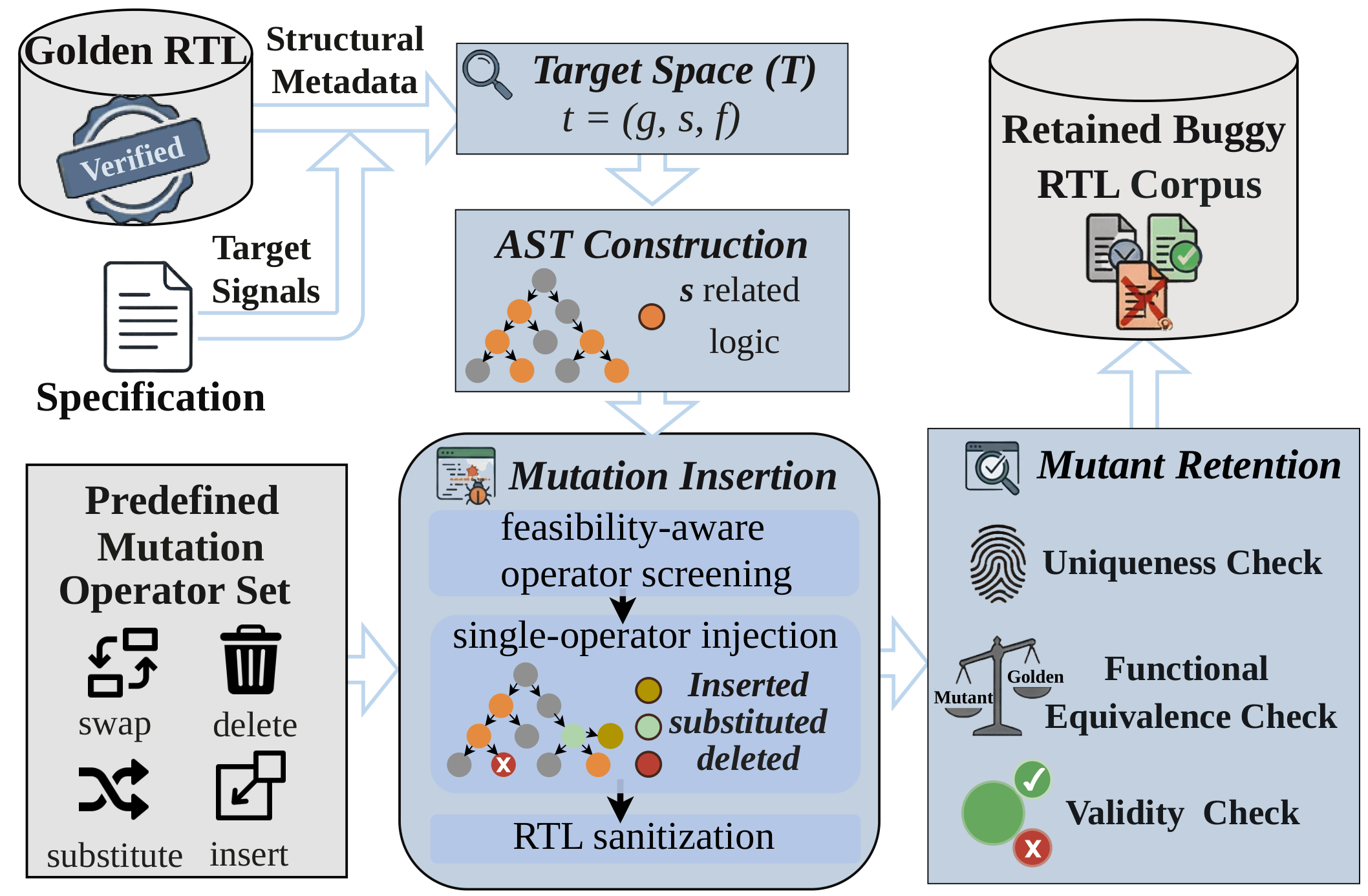}
\caption{Overview of the Mutation Insertion Methodology, detailing the pipeline from the verified golden RTL and structured specifications to the final retained buggy RTL corpus.}
\label{fig:mutation_flow}
\end{figure}

We generate buggy RTL variants through a constrained AST-level mutation workflow that balances behavioral relevance, structural validity, and diversity. 
As illustrated in \Cref{fig:mutation_flow}, the workflow proceeds from mutation-space definition to localized AST construction, feasibility-aware screening, and final mutant selection. 
Although AST-level mutations mainly capture localized faults, they provide a scalable lower-bound proxy for assertion effectiveness~\cite{wu2023mantra}.
We describe this workflow in detail below in three main steps.

% Step 1 2 3
\textbf{Step 1: Define mutation targets.} 
We begin by defining the mutation targets. 
Each target is specified by the resolved module, the target signal, and the corresponding RTL file.
Let $T$ denote the set of mutation targets. Each target $t \in T$ is represented as $(m, s, f)$, where $m$ is the resolved module, $s$ is the target signal, and $f$ is the corresponding RTL file.
To construct $T$, we trace specification-relevant signals upward through the hierarchy to their first non-trivial logic instantiation, bypassing purely structural pass-through wires. 
Clock and reset signals are explicitly filtered. 
This anchors mutation targets to documented and externally visible behaviors, rather than arbitrary internal states, and makes the evaluation fairer and more interpretable.

\textbf{Step 2: Construct feasible AST-level edits.} 
For each target, we apply feasibility-aware operator screening under a single-mutation-per-mutant setting. To reduce overhead, AST construction is limited to the resolved target module $m$. 
Candidate nodes are matched with applicable operators from \Cref{tab:OP} according to their local syntactic context, such as an \textit{if-condition} or \textit{module instantiation}.
Operator applicability is validated through probe execution on fresh AST clones, and only target-operator pairs that produce a concrete edit are retained. 
The resulting mutated RTL is then sanitized to correct known AST code-generation errors, such as declaration initializers that may otherwise be lowered into separate \texttt{assign} statements.  
A mutant is accepted only if the sanitized RTL can be reparsed and remains structurally consistent with the baseline design, except for the intended injected mutation.

\textbf{Step 3: Filter redundant, equivalent, and invalid mutants.}
Mutant retention then follows a staged filtering procedure for uniqueness, non-equivalence, and implementation validity. 
We first remove duplicate edits by checking whether the same mutation operation has already been applied to the same source line. 
We then perform functional equivalence checking (FEC) to eliminate equivalent mutants, ensuring that each retained mutant introduces a genuine functional deviation from the golden design. 
Finally, the remaining mutants are re-validated in JasperGold for parsing and elaboration correctness. 
Only mutants that pass this final validation step are retained in the buggy RTL corpus.

The surviving designs form the retained buggy RTL corpus. 
Since exhaustive enumeration is impractical, we adopt a budgeted generation strategy: for each design, we over-generate executable candidates by approximately $1.5 \times$ the target budget and prioritize distinct target signals over operator diversity. 
This yields a compact but behaviorally broader mutation corpus for downstream assertion evaluation.

AssertLLM2 directly provides 20 single-bug RTL and one five-bug RTL for each design used in our evaluation. We will also release the mutation workflow to enable users to generate additional buggy RTL variants for their own designs.

% we provide code also
\section{Benchmark Evaluation Framework}

Unlike prior work, which typically evaluates assertion generation in only a single scenario, our framework supports both \textit{bug-prevention} and \textit{bug-hunting} within a unified evaluation pipeline. 
We further adopt a complete and rigorous set of evaluation metrics to provide a systematic and comprehensive assessment beyond evaluations based solely on pass rate or structural coverage.
As shown in Fig.~\ref{fig:framework}, both settings follow the same core verification flow:  generated assertions are first checked for syntactic correctness, then verified on the reference RTL to assess provability and functional coverage, and finally applied to the buggy RTL corpus described in \Cref{subsec:mutate} to measure their effectiveness in detecting design errors. 
The difference lies in the task setup and final evaluation objective. 
In \textit{bug-prevention}, assertions are generated from the specification alone, and the evaluation focuses on whether they correctly capture the intended behavior of the design.
In \textit{bug-hunting}, the buggy RTL is additionally included in the input to evaluate its effectiveness in detecting design errors.
Specifically, we report four complementary metric groups:
\begin{itemize}
    \item \textbf{Syntax.} The proportion of generated assertion sets that are syntactically valid and cleanly elaborable.
    \item \textbf{FPV Outcomes.} Formal verification results on the golden RTL, reported as absolute counts of assertions classified as Proven, Counterexample (CEX), or Undetermined.
    \item \textbf{Multiple Coverage Metrics.} The extent of the verified design space, reported through COI coverage, proof coverage, and formal coverage.
    \item \textbf{Mutation-Based Bug Kill Ratio.} The fraction of valid, non-equivalent bugs killed by the generated assertions.
\end{itemize}

\begin{figure}[!t]
\centering
\includegraphics[width=0.9\linewidth]{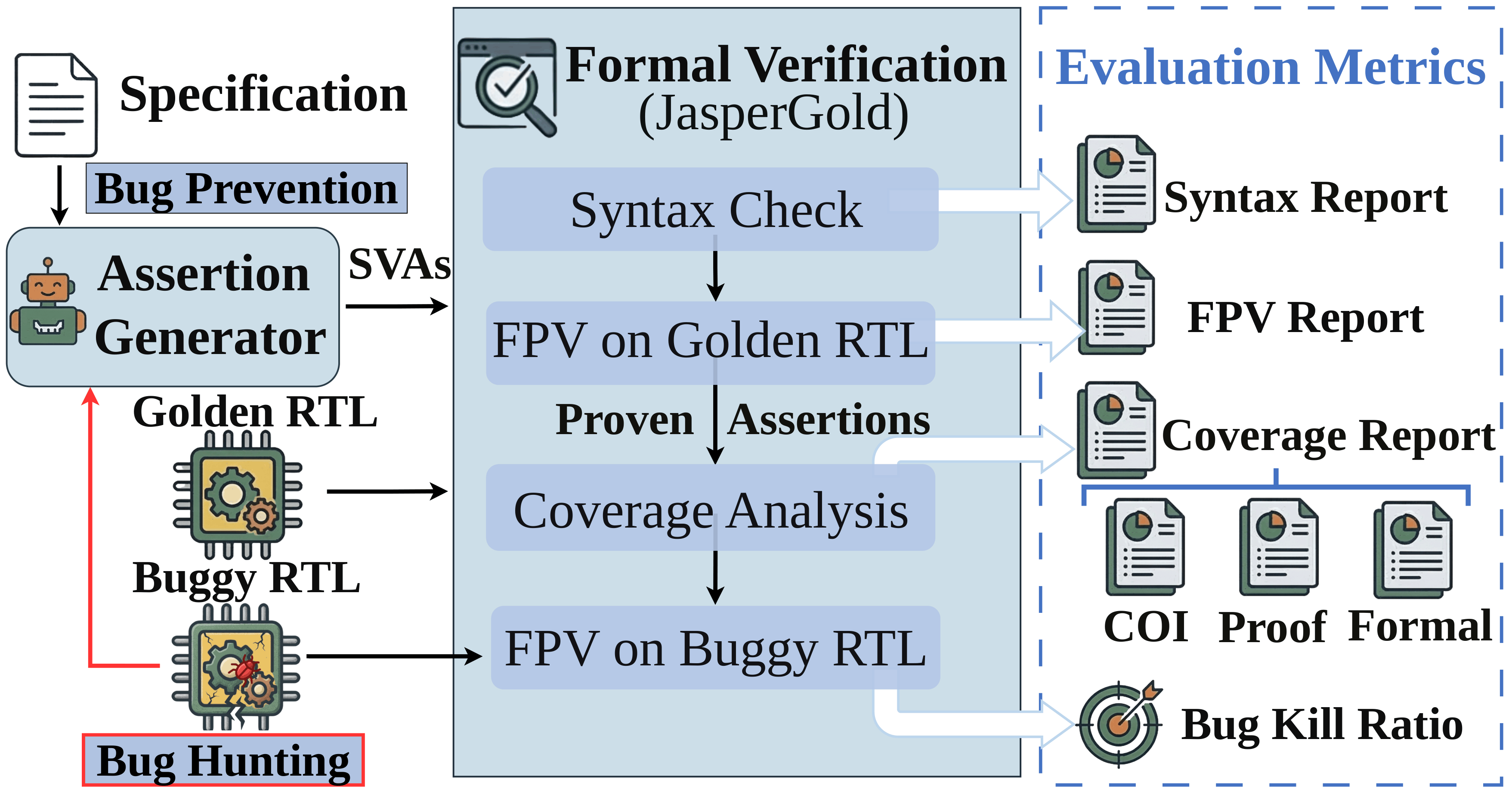}
\caption{Overview of the AssertLLM2 evaluation framework. Candidate SVAs are generated under either the bug-prevention or bug-hunting setting without access to the golden implementation. \looseness=-1}
\label{fig:framework}
\end{figure}

% We evaluate the generated assertions using four complementary metric groups:
% \begin{itemize}
%     \item \textbf{Syntax.} The proportion of generated assertion sets that are syntactically valid and cleanly elaborable.
%     \item \textbf{Coverage.} The extent of the verified design space, reported through COI coverage, proof coverage, and formal coverage.
%     \item \textbf{FPV Outcomes.} Formal verification results on the golden RTL, reported as absolute counts of assertions formally classified as Proven, Counterexample (CEX), or Undetermined.
%     \item \textbf{Mutation Kill Ratio (MKR).}The proportion of seeded functional mutants that are detected by the generated assertions.
% \end{itemize}

\paragraph{\textbf{Syntactic and Functional Validation.}}
Syntactic correctness is the baseline requirement. Generated SVAs are first parsed by the formal engine, and invalid assertions are discarded immediately. 
The remaining assertions are then checked by FPV against the golden RTL. 
A \textit{CEX} indicates that the assertion contradicts the design behavior, while an \textit{undetermined} result indicates that the solver does not complete the proof within the allotted budget. 
This stage, therefore, checks whether assertions are syntactically valid and behaviorally consistent with the golden RTL, but it does not assess their quality or practical verification value. 
Since many prior works focus primarily on these basic outcomes, they do not fully evaluate whether the generated assertions meaningfully capture functional intent.

\paragraph{\textbf{Coverage Analysis.}}
FPV outcomes alone do not fully characterize assertion quality. An assertion may be proven successfully while still being weak, shallow, or only loosely connected to the intended functionality. We therefore evaluate the proven assertions along 3 complementary coverage dimensions: cone-of-influence (COI) coverage, proof coverage, and formal RTL coverage.

% use coverage as the start
\textbf{COI coverage} provides a structural view of the upstream logic connected to the assertions. However, it reflects only structural reach and does not indicate whether that logic is meaningfully constrained.

\begin{figure}[!t]
\centering
\includegraphics[width=0.6\linewidth]{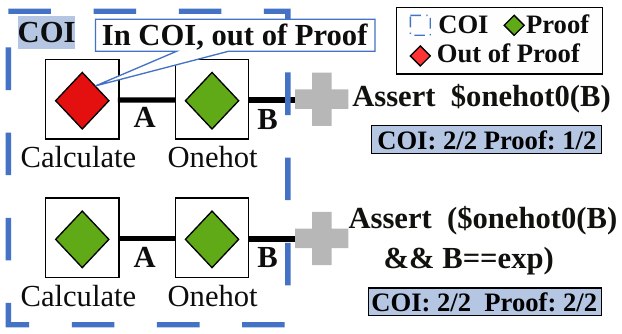}
\vspace{-0.3cm}
\caption{COI coverage versus proof coverage. A signal may lie in the structural COI of an assertion yet remain outside the proof core if the property constrains only shallow local behavior. Strengthening the property pulls more upstream logic into the proof core.}
\label{fig:COI}
\end{figure}

\textbf{Proof coverage} refines this view by identifying the subset of the COI that is actually required by the formal solver to prove the property. 
As illustrated in \Cref{fig:COI}, a shallow assertion such as \texttt{assert $onehot0(B)$} may have a large COI, while its proof coverage remains local. Only when the property depends on a deeper behavioral relation does more upstream logic enter the proof core. Proof coverage, therefore, provides a more precise measure of the logic substantively exercised by the assertions.

\textbf{Formal coverage} serves as a complementary RTL-level indicator that combines proof-related and stimuli-related information. 
Unlike proof-based views, it also reflects the convergence behavior of the FPV engine. 
It therefore provides an additional perspective on design-level coverage.
Since stimuli coverage is strongly affected by tool configuration rather than assertion quality, we use it only to validate the JasperGold setup and golden RTL environment, not as a primary benchmark metric.

% add the actual data

\paragraph{\textbf{Mutation-Based Bug Detection.}}
Even a proven assertion set with favorable coverage statistics does not necessarily indicate strong verification value. 
Assertions may still achieve nontrivial structural or proof coverage while checking only shallow properties, such as simple connectivity, width consistency, or localized interface behavior, without adequately capturing deeper functional intent. 
To evaluate whether generated assertions can genuinely expose meaningful design errors, we therefore incorporate mutation-based testing into the evaluation framework.

\begin{figure*}[t]
    \centering
        \centering
        \includegraphics[width=0.9\linewidth]{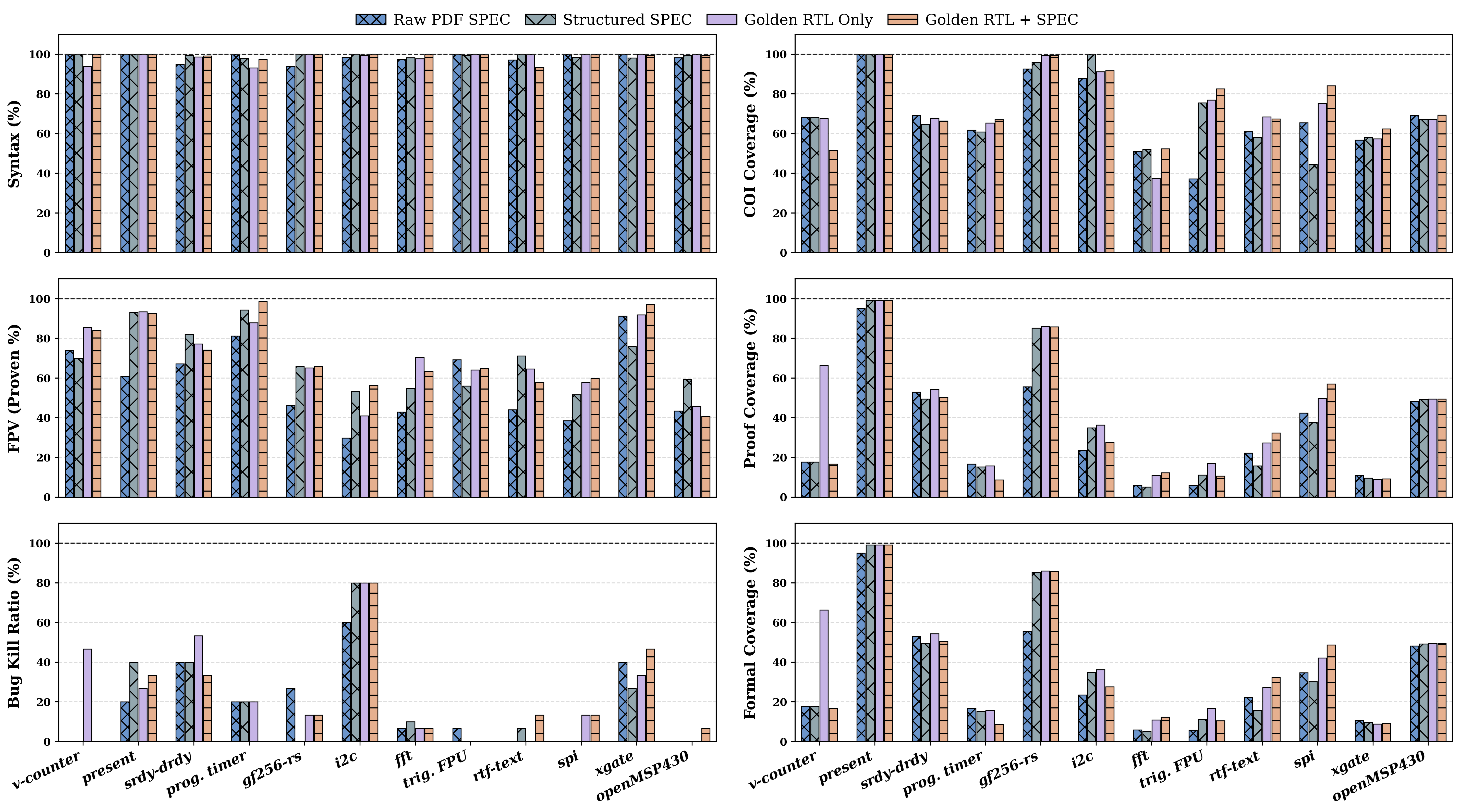}
    \vspace{-0.4cm}
    \caption{Per-design bug-prevention results across four generation settings on 12 representative designs.}
    \label{fig:bug_prevention}
\end{figure*}

Specifically, we apply the assertions proven on the golden RTL to the buggy RTL corpus introduced in \Cref{subsec:mutate}. 
For \textit{bug-prevention}, each design includes 20 single-bug buggy RTL variants. A variant is considered killed if at least one proven assertion is violated on it, and the \textbf{bug kill ratio} is defined as the fraction of these variants killed by the assertion set.
For \textit{bug-hunting}, each design includes one merged buggy RTL variant containing five injected bugs. In this case, the \textbf{bug kill ratio} is defined as the fraction of these five bugs killed by the assertion set.
It therefore complements syntax checking, FPV, and coverage analysis with a more direct measure of practical verification effectiveness.

\section{Experimental Results}

\subsection{Experimental Setup}

\subsubsection{Implementation and Models.}

\begin{table}[b]
\centering
\resizebox{\linewidth}{!}{%
\begin{tabular}{@{}l l c c c c c c@{}}
\toprule
\multirow{2}{*}{} 
& \multirow{2}{*}{\textbf{Model}}
& \multirow{2}{*}{\makecell[c]{\textbf{Syntax}\\\textbf{(\%)}}}
& \multirow{2}{*}{\makecell[c]{\textbf{FPV}\\\textbf{(P/T)}}}
& \multicolumn{3}{c}{\textbf{Coverage}}
& \multirow{2}{*}{\makecell[c]{\textbf{Bug Kill}\\\textbf{Ratio}}} \\
\cmidrule(lr){5-7}
&  &  &  & \textbf{COI} & \textbf{Proof} & \textbf{Formal} &  \\
\midrule

\multirow{6}{*}{Average}
& Gemini-2.5-Pro    & 88.29 & 13/20  & 62.47 & 20.19 & 19.16 &  9.76 \\
& Claude-Sonnet-4.5 & 93.04 & 21/38 & \textbf{\colorbox{lightred}{65.65}} & \textbf{\colorbox{lightred}{23.30}} & \textbf{\colorbox{lightred}{22.38}} & \textbf{\colorbox{lightred}{11.69}} \\
& GPT-5.2           & \textbf{\colorbox{lightred}{95.00}} & \textbf{\colorbox{lightred}{22/30}}  & 61.69 & 22.46 & 21.18 & 10.19 \\
& DeepSeek-V3.2     & 88.94 & 13/22  & 58.70 & 16.75 & 15.73 &  9.20 \\
& Qwen3-Coder-Plus  & 77.93 & 8/18  & 54.88 & 12.77 & 12.07 &  6.82 \\
& LLaMA-4-Maverick  & 89.23 & 5/13  & 37.27 &  9.09 &  8.46 &  5.08 \\
\midrule

\multirow{6}{*}{Union}
& Gemini-2.5-Pro    & 88.29 & 39/61  & 66.89 & 27.10 & 25.71 & \textbf{\colorbox{lightred}{18.51}} \\
& Claude-Sonnet-4.5 & 93.04 & 64/115 & \textbf{\colorbox{lightred}{68.36}} & \textbf{\colorbox{lightred}{29.72}} & \textbf{\colorbox{lightred}{28.32}} & 16.31 \\
& GPT-5.2           & \textbf{\colorbox{lightred}{95.00}} & \textbf{\colorbox{lightred}{65/91}}  & 64.06 & 28.62 & 27.15 & 14.87 \\
& DeepSeek-V3.2     & 88.94 & 38/65  & 61.51 & 25.66 & 24.17 & 15.32 \\
& Qwen3-Coder-Plus  & 77.93 & 25/53  & 55.96 & 19.72 & 18.55 & 12.18 \\
& LLaMA-4-Maverick  & 89.23 & 15/38  & 49.36 & 13.64 & 12.69 &  8.04 \\
\bottomrule
\end{tabular}%
}
\caption{Main results across different LLMs under the \textit{average} and the \textit{union} setting. \textit{P/T} denotes the number of proven assertions (\textit{P}) over the total number of syntactically correct assertions (\textit{T}) based on the FPV outcomes.}
\label{tab:main_results}
\end{table}

All experiments are conducted on a Linux workstation with an Intel Xeon 2.1\, GHz CPU and 512\, GB of memory. 
The benchmark pipeline is implemented in Python~3.11, with PyVerilog \cite{takamaeda2015pyverilog} used for AST parsing and buggy RTL generation. 
All evaluation is performed in Cadence JasperGold \cite{jaspergold2015}, while FEC is conducted using Cadence Conformal \cite{cadence_conformal_ug}.
We evaluate six state-of-the-art LLMs, including three closed-source commercial models (Gemini-2.5-Pro~\cite{comanici2025gemini}, Claude-Sonnet-4.5~\cite{claude_sonnet45_systemcard}, and GPT-5.2~\cite{singh2025openai}) and three leading open-weight models (DeepSeek-V3.2~\cite{liu2025deepseek}, Qwen3-Coder-Plus~\cite{Qwen3-Coder-Next}, and LLaMA-4-Maverick~\cite{llama4_maverick}).

\subsubsection{Evaluation Setting.}
% We evaluate the generated assertions using four complementary metric groups:
% \begin{itemize}
%     \item \textbf{Syntax.} The proportion of generated assertion sets that are syntactically valid and cleanly elaborable.
%     \item \textbf{Coverage.} The extent of the verified design space, reported through COI coverage, proof coverage, and formal coverage.
%     \item \textbf{FPV Outcomes.} Formal verification results on the golden RTL, reported as absolute counts of assertions formally classified as Proven, Counterexample (CEX), or Undetermined.
%     \item \textbf{Mutation Kill Ratio (MKR).}The proportion of seeded functional mutants that are detected by the generated assertions.
% \end{itemize}

% For bug-hunting, we inject five different bugs into each golden RTL and save the resulting designs as five separate single-bug mutants. 
% To ensure fair comparison across the two formulations, the same five single-bug mutants are also used as the downstream mutation corpus for the bug-prevention setting.

To account for generation variance, we run each model three times for every design. 
Results are reported in two forms: \textit{average} and \textit{union}. 
The \textit{average} result is the \textbf{default setting}, obtained by evaluating the three independently generated assertion sets separately and then taking the mean of their scores. 
This reflects typical single-run performance. 
The \textit{union} result serves as an additional view, obtained by merging the assertions from all three runs into a single set and evaluating that merged set in the same way as a standard output. 
This captures the cumulative value of multiple generations, as different runs may cover different parts of the design or expose different faulty variants. 
In the union setting, syntax and FPV proven rates remain the same as those in the \textit{average} setting, since both are calculated as overall proportions.

\subsection{Overall Benchmark Results.}
% report what and what is P/T
% do not compare the models, focus on the benchmark
\Cref{tab:main_results} shows the main evaluation results across different LLMs under \textit{average} and \textit{union} setting on all 83 designs in AssertLLM2.
For each model, we report syntax success rate, functional correctness as measured by the FPV proven-to-total ratio (\textit{P/T}),  COI coverage, proof coverage, formal RTL coverage, and bug kill ratio.

\Cref{tab:main_results} shows that surface-level correctness checks and structural coverage used in existing benchmarks are not sufficient to capture the main challenges of assertion generation. 
Most models already exceed 80\% syntax success, the strongest commercial models approach 95\%, their functional correctness exceeds 70\%, and even on the system-scale designs in AssertLLM2, leading models achieve over 60\% COI coverage. 
These results reveal a fundamental limitation of prior work: they often evaluate assertion generation only through easier metrics. Under these criteria, current models can appear highly capable. 
However, AssertLLM2 shows that strong syntax success, basic provability, and broad structural reach do not translate into equally strong proof coverage or bug detection capability. 
Proof coverage remains much lower than COI coverage, and the bug kill ratio is disproportionately low, peaking only in the high-20\% range even under the \textit{union} setting.
Although current LLMs are capable of producing well-formed and structurally relevant assertions, they still struggle to synthesize assertions that are consistently behaviorally rigorous and functionally discriminative.

We also observe a clear trade-off among the leading models. GPT-5.2 is more precision-oriented, generating fewer but more reliable assertions, whereas Claude-Sonnet-4.5 is more exploratory. 
Under prior metrics alone, GPT-5.2 would appear uniformly superior, and this difference in generation strategy would be largely invisible.

These findings show that AssertLLM2 provides a more challenging and realistic benchmark than prior evaluations and can therefore better guide future improvements in LLM-based assertion generation.

% introduce the setting, focus on the 5 bugs
\begin{figure}[t]
    \centering
        \centering
        \includegraphics[width=\linewidth]{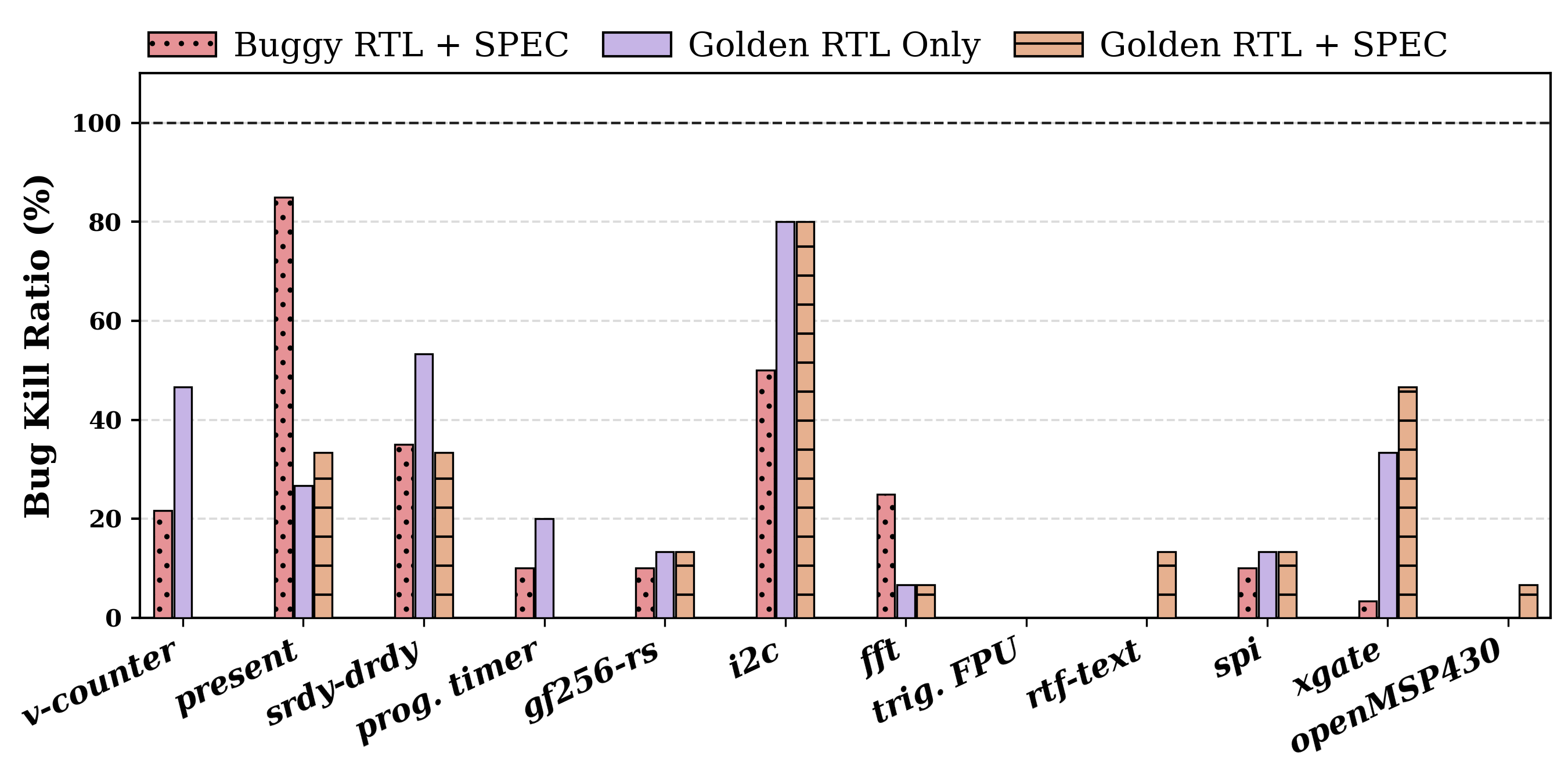}
    \vspace{-0.8cm}
    \caption{Per-design bug-hunting results, measured by bug kill ratio, across three generation settings on 12 designs.}
    \label{fig:bug_hunting}
\end{figure}

\subsection{Detailed Results on Selected Designs.}
To complement the aggregate results, we conduct a detailed design-level analysis of assertion generation under the two practical settings supported by AssertLLM2: \emph{bug-prevention} and \emph{bug-hunting} in \Cref{fig:bug_hunting,fig:bug_prevention}.
Owing to space constraints, we focus on 12 representative designs spanning diverse functions and scales. 
All results are reported under the default setting, averaged across three runs, using Claude Sonnet 4.5, the best-performing model overall.
For comparison, we also include two unrealistic golden-RTL-based settings adopted in prior work~\cite{pinckney2025cvdp, menon2025vert, pulavarthi2025assertionbench, kang2025fveval}. 
These settings are used only for reference, rather than as part of our benchmark settings.

\subsubsection{bug-prevention Detailed Results.} 
\Cref{fig:bug_prevention} presents the per-design \textit{bug-prevention} results on 12 representative designs across four generation settings, evaluated using syntax, FPV, proof coverage, formal coverage, COI coverage, and bug kill ratio.
The design-level variance in \Cref{fig:bug_prevention} demonstrates that assertion quality cannot be captured by a single metric.
High COI coverage, the only coverage metric used in prior work, does not guarantee strong proof coverage or effective bug-detection performance.
And better proof and formal coverage do not necessarily translate into better bug detection. 
For example, \textit{i2c} achieves high COI while its Proof coverage remains limited; \textit{PRESENT} shows only moderate mutation effectiveness despite near-saturated COI and Proof metrics; and \textit{xgate} attains a non-trivial but kill ratio despite low Proof coverage. 
These cases show that structural reach, formal provability, and bug kill capability are complementary but different dimensions of assertion quality. 
The key advantage of AssertLLM2 is therefore not only that it reveals such discrepancies, but also that it evaluates them jointly, leading to a more faithful and practically meaningful assessment. 
Omitting any of these dimensions risks overestimating model performance.

\Cref{fig:bug_prevention} also shows that input formulation has a clear impact on design-level performance. 
Raw specification introduces substantial variation across designs because the quality of the original documents is inconsistent. Some specifications are compact and behaviorally explicit, whereas others are fragmented, redundant, or uneven in detail. 
As a result, model performance becomes entangled with source-document quality, rather than reflecting only the ability to infer design intent. This effect is visible on the \textit{trigonometric FPU}, where Raw SPEC reduces average COI from 75.57\% to 37.17\% and average proof coverage from 11.17\% to 5.80\%. 
By contrast, structured specification provides a more stable and standardized specification interface, which makes cross-design evaluation more reliable. 
This is an important advantage of AssertLLM2, as it helps isolate model capability from variation in document quality.

% unrealistic
\subsubsection{bug-hunting Detailed Results.} \Cref{fig:bug_hunting} presents the per-design bug-hunting results on 12 representative designs across three generation settings, measured by bug kill ratio. For fairness, the two reference settings are also evaluated on the same five injected bugs in the buggy RTL.

Notably, \textit{Buggy RTL + SPEC}  significantly outperforms the idealized golden-RTL-based reference settings on some designs. 
For example, it substantially improves bug detection on \textit{PRESENT} and also outperforms both reference settings on \textit{fft}. 
This suggests that the gain does not come from a richer context alone, but from explicit fault-localized evidence in the buggy implementation. 
Such evidence can steer the model toward assertions that directly target the exposed discrepancy, rather than toward assertions that only restate general design intent. 

However, this benefit is not consistent across all designs. 
On larger designs such as \textit{openMSP430} and \textit{xgate}, \textit{Buggy RTL + SPEC} instead degrades performance substantially. 
As design complexity increases, faults become harder to localize, and the buggy RTL may introduce widespread behavioral evidence that conflicts with the specification. 
In such cases, the model can be misled into following incorrect implementation behavior rather than the intended design semantics. 

Overall, these results show that the bug-hunting setting in AssertLLM2 captures a practical capability omitted by prior benchmarks. 
Providing buggy RTL as input allows the benchmark to test whether a model can use spec--RTL inconsistencies to generate assertions that target exposed faults, while remaining robust when faults are less localized. 
In this way, AssertLLM2 enables a realistic evaluation that directly reflects practical bug detection capability and better aligns the evaluation with realistic verification workflows.

% \begin{figure}[t]
%     \centering
%         \centering
%         \includegraphics[width=\linewidth]{ICCAD25-Ada-Routing-v1/fig/SVA_complexity.png}
%     \vspace{-0.8cm}
%     \caption{Per-design comparison across 5 generation settings under the best-performing model using average with 3 runs. We select 12 representative designs spanning diverse categories for detailed analysis. The two golden-RTL-based settings are included only as idealized reference conditions for comparison, rather than as practical benchmark inputs.}
%     \label{fig:four_setting_comparison}
% \end{figure}

\section{Conclusion}
In this work, we presented \textbf{AssertLLM2}, an open-source benchmark for realistic SVA generation. 
AssertLLM2 addresses key limitations of prior work through a more complete task formulation, a more rigorous evaluation framework, and a richer benchmark construction spanning 83 real-world system-level designs. 
Our empirical study shows that, although current LLMs can often produce syntactically valid assertions, their practical verification value remains limited and highly uneven across designs and evaluation metrics.
By supporting both \textit{bug-prevention} and \textit{bug-hunting} within a unified benchmark, AssertLLM2 provides a stronger basis for evaluating model capability and advancing research toward practical automated hardware verification.

% \clearpage
\bibliographystyle{IEEEtran}
\bibliography{ref/Top-sim,ref/Assert}

\begin{thebibliography}{10}
\providecommand{\url}[1]{#1}
\csname url@samestyle\endcsname
\providecommand{\newblock}{\relax}
\providecommand{\bibinfo}[2]{#2}
\providecommand{\BIBentrySTDinterwordspacing}{\spaceskip=0pt\relax}
\providecommand{\BIBentryALTinterwordstretchfactor}{4}
\providecommand{\BIBentryALTinterwordspacing}{\spaceskip=\fontdimen2\font plus
\BIBentryALTinterwordstretchfactor\fontdimen3\font minus \fontdimen4\font\relax}
\providecommand{\BIBforeignlanguage}[2]{{%
\expandafter\ifx\csname l@#1\endcsname\relax
\typeout{** WARNING: IEEEtran.bst: No hyphenation pattern has been}%
\typeout{** loaded for the language `#1'. Using the pattern for}%
\typeout{** the default language instead.}%
\else
\language=\csname l@#1\endcsname
\fi
#2}}
\providecommand{\BIBdecl}{\relax}
\BIBdecl

\bibitem{liu2026llm}
H.~Liu, Y.~Lu, M.~Wang, X.~Yao, and B.~Yu, ``Llm-assisted circuit verification: A comprehensive survey,'' in \emph{2026 31st Asia and South Pacific Design Automation Conference (ASP-DAC)}, 2026.

\bibitem{witharana2022survey}
H.~Witharana, Y.~Lyu, S.~Charles, and P.~Mishra, ``A survey on assertion-based hardware verification,'' \emph{ACM Computing Surveys (CSUR)}, 2022.

\bibitem{foster2008assertion}
H.~Foster, ``Assertion-based verification: Industry myths to realities (invited tutorial),'' in \emph{International Conference on Computer Aided Verification}, 2008.

\bibitem{wu2025spec2assertion}
F.~Wu, E.~Pan, R.~Kande, M.~Quinn, A.~Tyagi, D.~K. Houngninou, J.~Rajendran, and J.~Hu, ``Spec2assertion: Automatic pre-rtl assertion generation using large language models with progressive regularization,'' \emph{arXiv preprint arXiv:2505.07995}, 2025.

\bibitem{fu2026chatsva}
L.~T. Fu, J.~Zhou, S.~Ren, M.~Zhang, J.~Xiong, H.~Jiang, N.~Guan, X.~Wang, and J.~Yang, ``Chatsva: Bridging sva generation for hardware verification via task-specific llms,'' \emph{arXiv preprint arXiv:2604.02811}, 2026.

\bibitem{tian2025assertcoder}
E.~Tian, Y.~Ci, Q.~Yang, Y.~Li, and Z.~Lyu, ``Assertcoder: Llm-based assertion generation via multimodal specification extraction,'' \emph{arXiv preprint arXiv:2507.10338}, 2025.

\bibitem{bai2025assertionforge}
Y.~Bai, G.~B. Hamad, S.~Suhaib, and H.~Ren, ``Assertionforge: Enhancing formal verification assertion generation with structured representation of specifications and rtl,'' in \emph{2025 IEEE International Conference on LLM-Aided Design (ICLAD)}, 2025.

\bibitem{lyu2025assertgen}
H.~Lyu, Y.~Wang, Y.~Du, M.~Shi, Z.~Chao, W.~Li, T.~Wang, and H.~Li, ``Assertgen: Enhancement of llm-aided assertion generation through cross-layer signal bridging,'' \emph{arXiv preprint arXiv:2509.23674}, 2025.

\bibitem{lyu2026assertminer}
H.~Lyu, Y.~Wang, J.~Zhou, Z.~Chao, T.~Wang, and H.~Li, ``Assertminer: Module-level spec generation and assertion mining using static analysis guided llms,'' in \emph{2026 31st Asia and South Pacific Design Automation Conference (ASP-DAC)}, 2026.

\bibitem{lyu2025assertfix}
H.~Lyu, Y.~Du, Y.~Wang, Z.~Chao, T.~Wang, and H.~Li, ``Assertfix: Empowering automated assertion fix via large language models,'' \emph{arXiv preprint arXiv:2509.23972}, 2025.

\bibitem{mali2024chiraag}
B.~Mali, K.~Maddala, V.~Gupta, S.~Reddy, C.~Karfa, and R.~Karri, ``Chiraag: Chatgpt informed rapid and automated assertion generation,'' in \emph{2024 IEEE Computer Society Annual Symposium on VLSI (ISVLSI)}, 2024.

\bibitem{shahidzadeh2025automated}
M.~Shahidzadeh, B.~Ghavami, S.~J. Wilton, and L.~Shannon, ``Automated verilog assertion generation using fine-tuned llms with subtask-specific iterative prompting,'' in \emph{2025 26th International Symposium on Quality Electronic Design (ISQED)}, 2025.

\bibitem{thangellamudi2026bridging}
J.~Thangellamudi and S.~M.~P. Dinakarrao, ``Bridging rtl and assertion generation with large language models,'' \emph{IEEE Design \& Test}, 2026.

\bibitem{menon2025vert}
A.~Menon, S.~Miftah, S.~Kundu, S.~Kundu, A.~Srivastava, A.~Raha, G.~Sonnenschien, S.~Banerjee, D.~Mathaikutty, and K.~Basu, ``Enhancing large language models for hardware verification: A novel systemverilog assertion dataset,'' \emph{ACM Transactions on Design Automation of Electronic Systems}, 2025.

\bibitem{pulavarthi2025assertionbench}
V.~Pulavarthi, D.~Nandal, S.~Dan, and D.~Pal, ``Assertionbench: A benchmark to evaluate large-language models for assertion generation,'' in \emph{Findings of the Association for Computational Linguistics: NAACL 2025}, 2025.

\bibitem{kang2025fveval}
M.~Kang, M.~Liu, G.~B. Hamad, S.~M. Suhaib, and H.~Ren, ``Fveval: Understanding language model capabilities in formal verification of digital hardware,'' in \emph{2025 Design, Automation \& Test in Europe Conference (DATE)}, 2025.

\bibitem{pinckney2025cvdp}
N.~Pinckney, C.~Deng, C.-T. Ho, Y.-D. Tsai, M.~Liu, W.~Zhou, B.~Khailany, and H.~Ren, ``Comprehensive verilog design problems: A next-generation benchmark dataset for evaluating large language models and agents on rtl design and verification,'' \emph{arXiv preprint arXiv:2506.14074}, 2025.

\bibitem{fang2024assertllm}
W.~Fang, M.~Li, M.~Li, Z.~Yan, S.~Liu, Z.~Xie, and H.~Zhang, ``Assertllm: Generating and evaluating hardware verification assertions from design specifications via multi-llms,'' \emph{arXiv preprint arXiv:2402.00386}, 2024.

\bibitem{alshazly2014detecting}
A.~A. Alshazly, A.~M. Elfatatry, and M.~S. Abougabal, ``Detecting defects in software requirements specification,'' \emph{Alexandria Engineering Journal}, 2014.

\bibitem{chang2024natural}
K.~Chang, Z.~Chen, Y.~Zhou, W.~Zhu, K.~Wang, H.~Xu, C.~Li, M.~Wang, S.~Liang, H.~Li \emph{et~al.}, ``Natural language is not enough: Benchmarking multi-modal generative ai for verilog generation,'' in \emph{Proceedings of the 43rd IEEE/ACM International Conference on Computer-Aided Design}, 2024.

\bibitem{freecores}
{FreeCores}, ``{FreeCores},'' \url{https://github.com/orgs/freecores}.

\bibitem{e203_hbirdv2}
{riscv-mcu}, ``{Hummingbirdv2 E203 Core and SoC},'' \url{https://github.com/riscv-mcu/e203_hbirdv2}.

\bibitem{opencores}
{OpenCores}, ``{OpenCores},'' \url{https://opencores.org/}.

\bibitem{vexriscv}
{SpinalHDL}, ``{VexRiscv: A FPGA friendly 32 bit RISC-V CPU implementation},'' \url{https://github.com/SpinalHDL/VexRiscv}.

\bibitem{jaspergold2015}
{Cadence Design Systems, Inc.}, \emph{{JasperGold Apps User's Guide}}, Cadence Design Systems, Inc., Sep. 2015.

\bibitem{wu2023mantra}
J.~Wu, Y.~Lei, Z.~Zhang, X.~Meng, D.~Yang, P.~Li, J.~He, and X.~Mao, ``Mantra: Mutation testing of hardware design code based on real bugs,'' in \emph{2023 60th ACM/IEEE Design Automation Conference (DAC)}, 2023.

\bibitem{takamaeda2015pyverilog}
S.~Takamaeda-Yamazaki, ``Pyverilog: A python-based hardware design processing toolkit for verilog hdl,'' in \emph{International Symposium on Applied Reconfigurable Computing}, 2015.

\bibitem{cadence_conformal_ug}
{Cadence Design Systems, Inc.}, \emph{Conformal Equivalence Checking User Guide}, Cadence Design Systems, Inc., Oct. 2021.

\bibitem{comanici2025gemini}
G.~Comanici, E.~Bieber, M.~Schaekermann, I.~Pasupat, N.~Sachdeva, I.~Dhillon, M.~Blistein, O.~Ram, D.~Zhang, E.~Rosen \emph{et~al.}, ``Gemini 2.5: Pushing the frontier with advanced reasoning, multimodality, long context, and next generation agentic capabilities,'' \emph{arXiv preprint arXiv:2507.06261}, 2025.

\bibitem{claude_sonnet45_systemcard}
{Anthropic}, ``Claude sonnet 4.5 system card,'' Anthropic, Tech. Rep., 2025.

\bibitem{singh2025openai}
A.~Singh, A.~Fry, A.~Perelman, A.~Tart, A.~Ganesh, El-Kishky \emph{et~al.}, ``Openai gpt-5 system card,'' \emph{arXiv preprint arXiv:2601.03267}, 2025.

\bibitem{liu2025deepseek}
A.~Liu, A.~Mei, B.~Lin, B.~Xue, B.~Wang, B.~Xu, B.~Wu, B.~Zhang, C.~Lin, C.~Dong \emph{et~al.}, ``Deepseek-v3.2: Pushing the frontier of open large language models,'' \emph{arXiv preprint arXiv:2512.02556}, 2025.

\bibitem{Qwen3-Coder-Next}
R.~Cao, M.~Chen, J.~Chen, Z.~Cui, Y.~Feng, B.~Hui, Y.~Jing, K.~Li, M.~Li, J.~Lin \emph{et~al.}, ``Qwen3-coder-next technical report,'' \emph{arXiv preprint arXiv:2603.00729}, 2026.

\bibitem{llama4_maverick}
\BIBentryALTinterwordspacing
{Meta}, ``{The Llama 4 herd: The beginning of a new era of natively multimodal AI innovation}.'' [Online]. Available: \url{https://ai.meta.com/blog/llama-4-multimodal-intelligence}
\BIBentrySTDinterwordspacing

\end{thebibliography}

\end{document}